\documentclass[12pt,letterpaper]{article}

\usepackage{fixltx2e}
\usepackage{textcomp}
\usepackage{fullpage}
\usepackage{amsfonts}
\usepackage{verbatim}
\usepackage[english]{babel}
\usepackage{pifont}
\usepackage{color}
\usepackage{setspace}
\usepackage{lscape}
\usepackage{indentfirst}
\usepackage[normalem]{ulem}
\usepackage{booktabs}
\usepackage{float}
\usepackage{latexsym}
\usepackage{url}
\usepackage{hyperref}
\usepackage{epsfig}
\usepackage{graphicx}
\usepackage{amssymb}
\usepackage{amsmath}
\usepackage{bm}
\usepackage{array}
\usepackage[version=3]{mhchem}
\usepackage{ifthen}
\usepackage{caption}
\usepackage{hyperref}
\usepackage{amstext}

\newtheorem{theorem}{Theorem}

\newtheorem{lemma}{Lemma}

\renewcommand{\section}[1]{%
\bigskip
\begin{center}
\begin{Large}
\normalfont\scshape #1
\medskip
\end{Large}
\end{center}}

\renewcommand{\subsection}[1]{%
\bigskip
\begin{center}
\begin{large}
\normalfont\itshape #1
\end{large}
\end{center}}

\renewcommand{\subsubsection}[1]{%
\vspace{2ex}
\noindent
\textit{#1.}---}

\renewcommand{\tableofcontents}{}

\begin{document}
\noindent RH:  Tree-Based Phylogenetic Networks

\bigskip
\medskip
\begin{center}

\noindent{\Large \bf On Tree Based Phylogenetic Networks}
\bigskip
\vspace{1.5in}


\noindent {\normalsize \sc Louxin Zhang$^1$
}\\
\noindent {\small \it 
$^1$Department of Mathematics, National University of Singapore, Singapore 119076 
}\\
\end{center}
\medskip
\vspace{1in}

{\bf Corresponding author:} \\
\begin{center}
LX Zhang\\
Department of Mathematics\\
National University of Singapore\\
Singapore 119076\\
Tel: +65-65166579\\
 E-mail: matzlx@nus.edu.sg.
\end{center}

\newpage 
\subsubsection{Abstract}
A large class of phylogenetic networks can be obtained from trees by the addition of horizontal edges between the tree edges. These networks are called tree based networks. Reticulation-visible networks and child-sibling networks are all tree based. In this work, we present
a simply necessary and sufficient condition for tree-based networks and prove that there is a universal tree based network $N$ for each set of species such that every phylogenetic tree on the same species is a base of $N$.  The existence of universal tree based network implies that for any given set of phylogenetic trees (resp. clusters) on the same species there exists a tree base network
that display all of them.
\vspace{1em}

\noindent {\bf Keywords}: Phylogenetic tree, phylogenetic network, tree base, reticulation visibility,, horizontal gene transfer\\


\vspace{0.5in}

Reticulation process refers to the transfer of genes between organisms in a way other than reproduction. 
One of the major reticulation processes is horizontal gene transfer.  It has been considered as a highly significant   form of genetic transfer among single-cell organisms (Doolittle, 1999; Doolittle and Bapteste, 2007; Smets and Barkay, 2006; Treangen and  Rocha, 2011). Other reticulation processes include introgression,  recombination and hybridization (Dagan and Martin, 2006; 
Fontaine et al., 2015;  McBreen and Lockhart, 2006). 

A set of gene trees are usually reconciled into a phylogenetic network to model reticulation processes (Doolittle and Bapteste, 2007; Huson, Rupp, and Scornavacca, 2011).  A phylogenetic network is a rooted acyclic  digraph in which there is a special node of out-degree 2 and in-degree 0 (called the root) such that all the edges are directed away from it and
the set of in-degree 0 nodes correspond one-to-one the collection of present-day taxa under study. 
A network is binary if every node other than the root and leaves is of degree three. 
Clearly, a phylogenetic tree is a binary phylogenetic network without reticulation nodes. 

Horizontal gene transfers are naturally modeled and visualized by using a tree-based phylogenetic network, where the underlying base tree represents the evolution of the species from which genes are sampled and  branches  are added between  tree branches to represent  horizontal gene transfers (Smets and Barkay, 2006; Nakhleh, 2013).  Surprisingly,  phylogenetic networks that are used for modeling other reticulation processes may also have the same topological structure,  obtained from a tree by the addition of branches between tree branches, even if not every binary phylogenetic network shares this property (van Iersel, 2013).    
Recently, Francis and Steel (2015) initiated the study of  tree-based networks. In their paper,  sufficient conditions for tree-based networks are presented. They further showed that
this class of networks include reticulation visible networks and tree-sibling networks. 

In the present work, we answer two problems posed by Francis and Steel (2015). Precisely, we present a simple necessary and sufficient condition for tree-based networks. We also construct a universal network on $X$  that has every tree on $X$ as its base for any $X$ of an arbitrary size.

\section{Tree-based Networks}

\subsection{Basic Definitions}
A digraph $D$ consists of a set of vertexes, $V(D)$,  and a collection of directed edges,  $E(D)$, that each connects an ordered pair of vertexes.
We call $(u, v)\in E(D)$ an outgoing edge of $u$ and an incoming edge of $v$.  
For each $x\in V(D)$, 
 the number of the incoming edges  of $x$  is called its \emph{indegree}; the number of the outgoing edges of $x$  is
call its \emph{outdegree}; the sum of the indegree and outdegree of $x$ is  called its \emph{degree}. The indegree, outdegree, and degree of   $x$
are writeen  $d^{i}(x)$, $d^{o}(x)$ and $d(x)$, respectively.

A path from $x$ to $y$ in $D$ is made up of two or more ``successive" vertexes
$x=u_1, u_2, \cdots, u_k=y$,
where  $(u_i, u_{i+1})\in E(D)$ for
$1\leq i\leq k-1$ and $k\geq 2$. 
%
 A cycle is a path from a node to itself. 
 $D$ is acyclic if it does not contain any cycle.

A binary phylogenetic network  over a set $X$ of species is an 
acyclic digraph with the following properties: 
\begin{itemize}
\item There exists a unique vertex $\rho$  such that  $d^{i}(\rho)=0$. It  is  the
\emph{root}  of the network.  The root is of outdegree 2.
\item There are exactly $|X|$ nodes $\ell$ such that $d^{i}(\ell)=1$ and $d^{o}(\ell)=0$,  corresponding one-to-one with the species. These nodes are called the \emph{ leaves}  of the network. 
\item All the vertexes that are neither a leaf nor the root are of degree three. They are called {\it internal nodes}.
\end{itemize} 

An internal node $x$ in a binary phylogenetic network  is called a {\it tree} (or {\it speciation}) {\it node} if $d^{i}(x)=1$ and $d^{o}(x)=2$; it is called a {\it reticulation} {\it node} if  $d^{i}(x)=2$ and $d^{o}(x)=1$.  Since the root is the only vertex having indegree 0 in a phylogenetic network, there is a path from the root to every other vertex.  For two vertexes $x$ and $y$,  if there is a path from $y$ to $x$, $y$ is said to be an \emph{ancestor} of $x$ and $x$ is said to be a \emph{descendant} of $y$.

\begin{figure}[!t]
\centering
\includegraphics[width=0.5\textwidth]{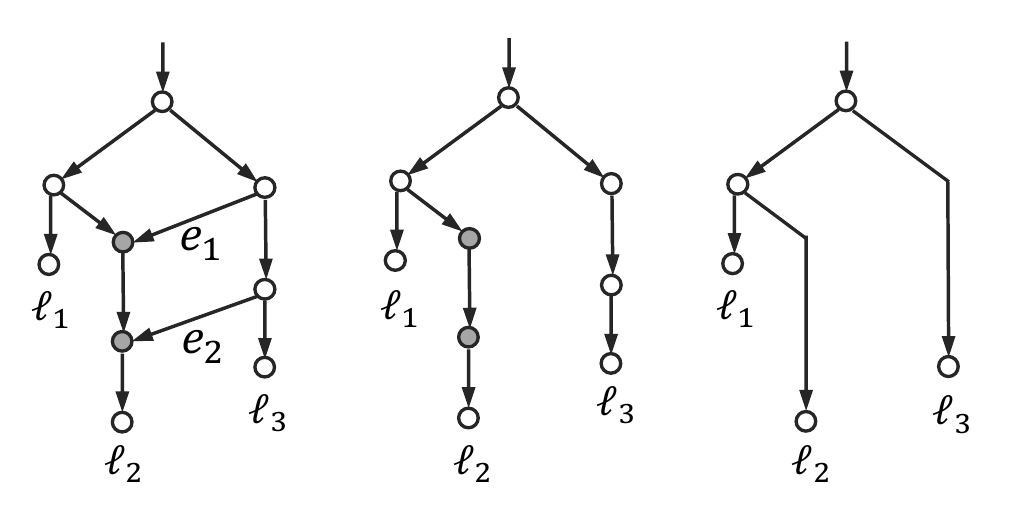}
\caption{A tree-based phylogenetic network (left) and a tree base of it (right). The subdivision of the base tree (middle) is a subtree of the network that can be obtained by the removal of the edges $e_1$ and $e_2$.  Reticulation nodes in the network are represented by shaded circles,
\label{fig1}}
\end{figure}

A binary phylogenetic networks is shown in Figure~\ref{fig1}, where we draw an open branch entering the root, representing the least common ancestor of all the species. 
In rest of the paper,  for a binary phylogenetic network $N$,   we shall use the following notation: 
\begin{itemize}
\item $\rho_N$: The root of $N$;\vspace{-0.5em}
\item ${\cal V}(N)$: The set of nodes in $N$; \vspace{-0.5em}
\item ${\cal T}(N)$:  The set of tree nodes in $N$;\vspace{-0.5em}
\item ${\cal  R}(N)$: The set of reticulation nodes in $N$;\vspace{-0.5em}
\item ${\cal E}(N)$: The set of (directed) edges in $N$; \vspace{-0.5em}
\item ${\cal L}(N)$: The set of labelled leaves in $N$; \vspace{-0.5em}
\item ${ c}(u)$: The unique child of $u$ if $u\in {\cal  R}(N)$, or the set of the children of $u$ if $u \in {\cal T}(N)$;\vspace{-0.5em}
\item $p(u)$: The unique parent of $u$ if $u\in {\cal T}(N)$, or the set of the parents of $u$ if $u \in {\cal R}(N)$;
\end{itemize}

\subsection{Tree-based networks}

Let $N$ be a network over a set of species, $X$.
For a subset  $E\subseteq {\cal E}(N)$, $N-E$ denotes the subnetwork of $N$ obtained after the removal of the edges in $E$. 
If $E$ contains exactly an incoming edge for each reticulation vertex, then 
  every non-root node in $N-E$ is of indegree 1 and hence is a tree.  However, it may contain new leaves.  $N$ is {\it tree-based}  if there exists $E\subseteq {\cal E}(N)$ such that 
$N-E$ is a subtree having the same leaves as $N$.

The network in Figure~\ref{fig1}
is tree-based.  It has two reticulation vertexes.
The edge $e_1$ enters the top reticulation vertex, whereas $e_2$ is an edge entering the other  at the bottom.  The removal of these two edges 
results in a subtree with the same
leaves as the network. 
On the other hand, 
the network in Figure~\ref{fig2}A is not 
tree-based. The reason is that no matter which of  the incoming edges
$(r_1, r_4)$ and $(r_2, r_4)$ is removed for $r_4$, the tail of the removed edge becomes a new leaf in the resulting subtree.

Tree-based networks compose of a large class of interesting networks.  A vertex in a phylogenetic network is called 
\emph{visible} (or stable)  if there exists a leaf such that every path from the network root to the leaf passes through the vertex. 
A network is \emph{reticulation visible} if every reticulation vertex is visible.  Reticulation visible networks are tree-based (Francis and Steel, 2015; Gambette {\it et al.}, 2015).  

A phylogenetic network is \emph{tree sibling} 
if every reticulation vertex has a tree vertex sibling. Tree sibling networks are also tree based (Francis and Steel, 2015).

\newpage

\section{Main Results} 


\subsection{A necessary and sufficient condition for tree-based networks}

\begin{figure}[!t]
\centering
\includegraphics[width=0.6\textwidth]{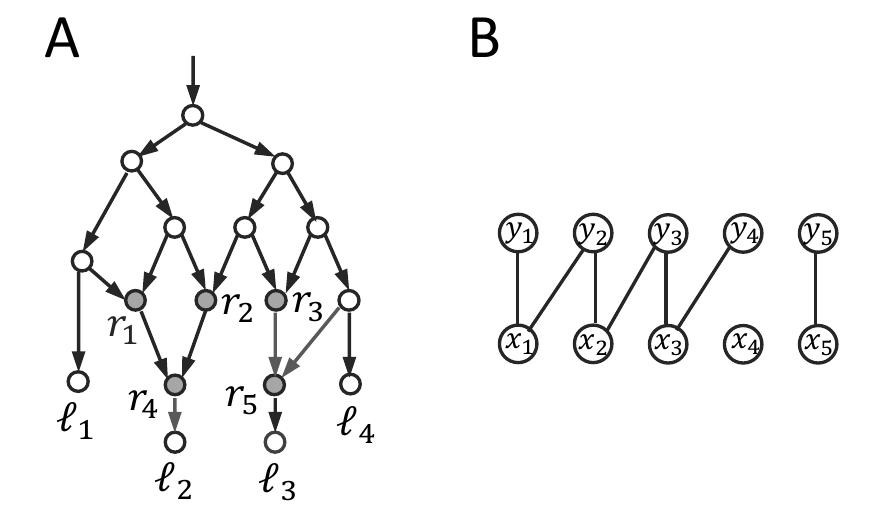}
\caption{({\bf A}) A binary phylogenetic network $N$, in which reticulation vertexes are represented by shaded circles. ({\bf B}) The bipartite graph $B(N)$ defined for $N$.
\label{fig2}}
\end{figure}

In a binary phylogenetic  network, a reticulation vertex is said to be of:
\begin{itemize}
\item  {\it type-0} if  if its parents are both a reticulation vertex;
\item  {\it type-1} if a parent is a reticulation vertex and the other is a tree vertex;
\item {\it type-2}  if its parents are both a tree vertex.
\end{itemize}
In the network drawn in Fig~\ref{fig2}A., the vertexes
$r_1, r_2$ and $r_3$ are of type-2, $r_5$ is of type-1, and $r_4$ is of type-0. 
A tree-based network must not contain any type-0 reticulation vertexes (Francis and Steel, 2015). 

Let $N$ be a binary phylogenetic network without type-2 vertexes.  Setting
${\cal R}(N)=\{r_1, r_2, ..., r_s\}$, 
we define an undirected bipartite graph $B(N)=(X\cup Y, E)$ as follows: 
\begin{eqnarray*}
  &&X=\{x_1, x_2, \ldots, x_{s}\;|\; x_i \mbox{ represents  $r_i$ for each $i$} \}, \\
  &&Y=\{y_1, y_2, \ldots, y_{t}\;|\; y_i \mbox{ represents a  parent  in ${\cal T}(N)$  
 of a
  vertex  in ${\cal R}(N)$ }\}, 
\end{eqnarray*} 
and 
\begin{eqnarray*}
 E = \{ (y_j, x_i) \;|\;  \mbox{the vertex represented by $y_j$ is a parent of the vertex by $x_i$ in $N$.} \}. 
\end{eqnarray*} 
Remark that $B(N)$ is essentially a bipartite subgraph of 
$N$.  For example, Figure~\ref{fig2}B shows the bipartite network defined for the network in Figure~\ref{fig2}A, 
in which $x_4$ is not connected with any other vertex, as the parents of $r_4$ are both not a tree vertex.

Using the technique of Gambette {et al.} (2015), we are able to present a
simple  necessary and sufficient condition for  binary tree-based phylogenetic  networks.  

\begin{lemma}
\label{tree_base_theorem}
Let $N$ be a network without type-0 reticulation vertexes. Then $N$ is tree-based if and only if for every  two type-1 reticulation vertexes, their  correspondences are not connected in $B_N$.   
\end{lemma} 
\noindent {\bf Proof.} 
 First, we have the following two facts:
 
(i) Let  $e=(x, y)\in {\cal E}(N)$. 
If $x$ is a reticulation vertex, then $x$ has out-degree 0 and hence becomes  a leaf in  $N-\{e\}$.

(ii) Let $e_1=(x_1, y_1)\in {\cal E}(N)$ and $e_2=(x_2, y_2)\in {\cal E}(N)$ such that $x_i\in {\cal T}(N)$ and $y_i\in {\cal R}(N)$ for $i=1, 2$. If $x_1=x_2$, then $x_1$ becomes a leaf in $N-\{e_1, e_2\}$. If $y_1=y_2$, then $y_1$ has in-degree 0 in $N-\{e_1, e_2\}$. 

Each edge $(t, r)$ in $N$ corresponds an edge in $B(N)$, where $t\in {\cal T}(N)$ and $r\in {\cal R}(N)$.  For a subset $E\subseteq {\cal E}(N)\cap \left( {\cal T}(N)\times {\cal R}(N)\right)$, we set 
$E_{B(N)}$ to be the subset of edges  in $B(N)$ that  correspond one-to-one  to the edges in $E$. 
The two facts stated above imply that $N-E$ is a tree network with the same leaves as $N$ if and only if 
$E$ is a matching covering every reticulation vertex in $N$ and hence if and only if $E_{B(N)}$ is a complete matching from $X$ to $Y$ in $B(N)$. 

Since $B(N)$ is bipartite, by Hall's theorem,  
there is a complete matching from $X$ to $Y$ if and only if 
 $|X'|\leq |N(X')|$ for any $X'\subseteq X$, where $N(X')$ is the set of vertexes that are adjacent with some vertexes in $X'$, and so if and only if 
 there is a complete matching from $C\cap X$ to $C\cap Y$ for every connected component $C$ in $B(N)$. 
 
 A vertex $x_i$ in $B(N)$ is of degree 1 if it corresponds a type-1 reticulation vertex; it is of degree 2 if it corresponds a type-2  reticulation vertex.  Each vertex $y$ in $B(N)$ has also  degree 1 or degree 2, as the tree vertex represented  $y$ has one or two reticulation children.
 Therefore,  every connected component is either a cycle or a path in $B(N)$. 
 Let $C$ be a connected component in $B(N)$. If $C$ is a cycle, 
 $C$ has a perfect matching from $C\cap X$ to $C\cap Y$. If $C$ is a path, it contains exactly 
 two degree-1 vertexes $w'$ and $w''$.  There is a complete matching from $C\cap X$ to $C\cap Y$ if and only if either $w'$ or $w''$ is not in $X$. 
 
 Since the degree-1 vertexes  in $X$ correspond one-to-one to the type-1 reticulation vertexes,  we conclude that $N$ is tree-based if and only if 
the correspondences of every two type-1  reticulation vertexes 
are not connected in $B(N)$. 
$\Box$
\vspace{1em} 

\begin{figure}[!t]
\centering
\includegraphics[width=0.6\textwidth]{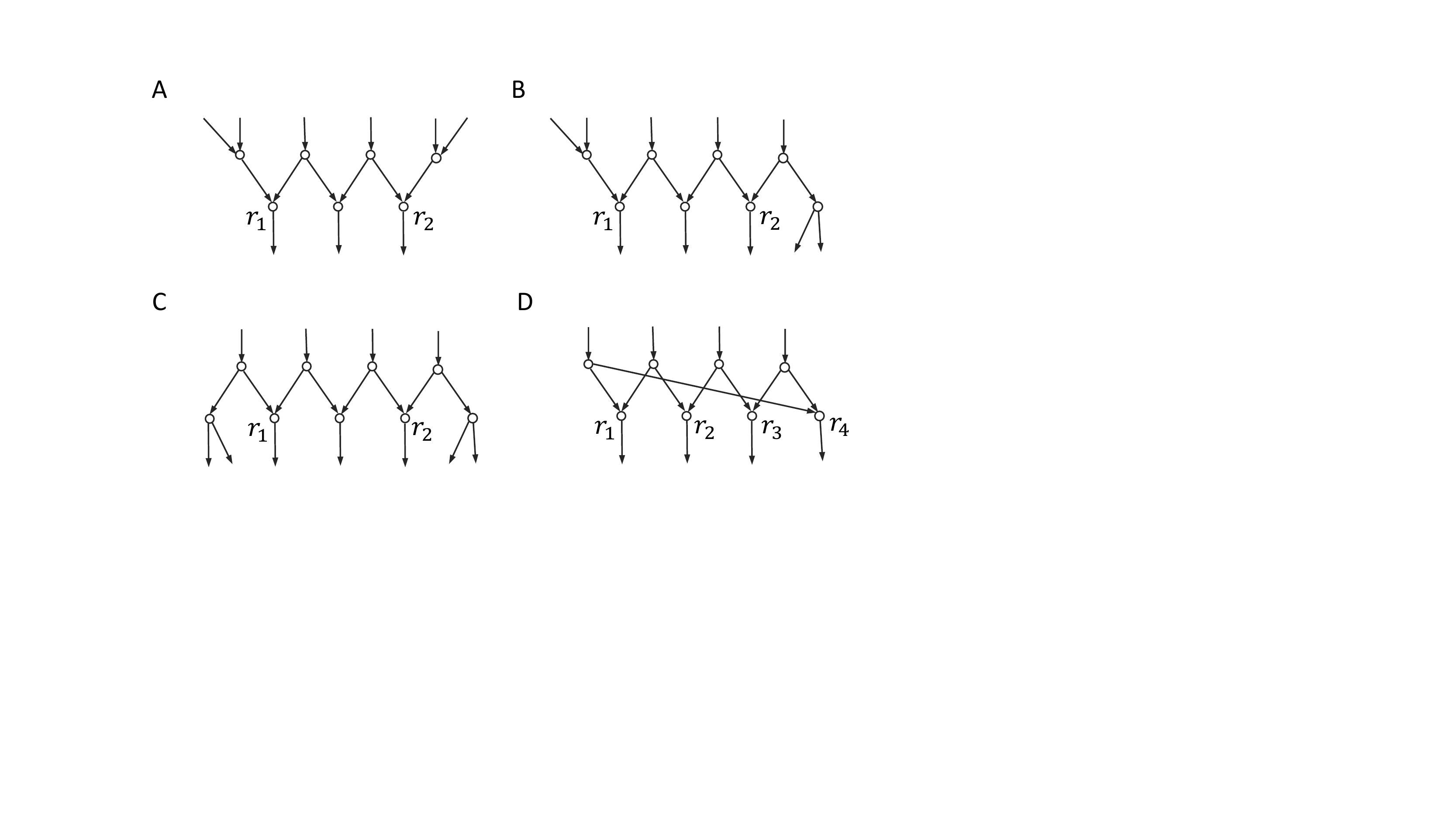}
\caption{({\bf A}) A maximal zigzag path between two type-1  reticulation vertexes $r_1$ and $r_2$. ({\bf B}) A maximal zigzagy path  between a type-1 reticulation vertex $r_1$ and a type-2 reticulation  vertex $r_2$.
({\bf C}) A maximal zigzagy path between two type-2 reticulaiton  vertexes $r_1$ and $r_2$.
({\bf D}) A zigzagy cycle consisting of only type-2 reticulation vertexes.}
\label{fig2_ZigZag}
\end{figure}

Let $u, v\in {\cal R}(N)$. We say that they are  connected by a {\it zigzagy}  path  if there is a sequence of vertexes $u=x_0, x_1, \ldots, x_{2k}=v$ such that the vertexes alternate between reticulation vertexes and their  tree vertex parents (Figure~\ref{fig2_ZigZag}). 

Recalled that 
 $B(N)$ is a disjoint union of paths and cycles.  Obviously, each cycle contains  only type-2 reticulation vertexes. Each type-1 reticulation vertex appears only  at the  ends of a path. Therefore, by Lemma~\ref{tree_base_theorem}, we have the following theorem.

\begin{theorem}
\label{tree_base_theorem2}
Let $N$ be a binary network. $N$ is tree-based if and only if
{\rm (i)} there is no type-0 reticulation vertex in $N$, and 
{\rm (i)} no two type-1 vertexes are connected by a zigzagy path. 
\end{theorem} 

Theorem~\ref{tree_base_theorem2} implies the following  algorithm for determining  whether
a network is tree-based or not. 

\begin{center}
\begin{tabular}{l} 
 \hline  
{\bf Input} A binary network $N$;\\
\\
1.  {\bf If}  it contains a 
type-0 reticulation vertex, {\bf output} ``$N$ is not tree-based"; \\
2.  {\bf Do} until there is no unmarked type-1 reticulation vertex $\{$\\
\hspace{2em} Select an unmarked type-1 reticulation vertex $u$;\\  
\hspace{2em} {\bf If}  the zigzagy path starting at $u$ terminates at an \\
\hspace{3.5em} unmarked type-1 reticulation vertex, {\bf output} ``$N$ is not tree-based";\\
\hspace{2em}  {\bf else}  mark $u$;\\ 
3. {\bf Output} ``$N$ is tree-based";\\
\hline
\end{tabular}
\end{center}

Obviously, the above algorithm is correct. Since any two zigzagy paths are disjoint, it takes a linear time. 

\section{Universal tree-based networks}


It is known that
there exists a network that displays every phylogenetic tree on the same species (see, for example, Francis and Steel, 2015). However,  a tree may be displayed, but not as a base,  in a phylogenetic network. 
Therefore, the following question is posed by Francis and Steel:
  \begin{quote}
    Does there exist a network $U$ over $X$ such that 
     every phylogenetic tree over $X$  is a base for $U$ for every large set of species $X$?
  \end{quote}
For $|X|=3$, such a universal network exists (Francis and Steel, 2015).  We shall present  such a universal tree-based network $U$ for every $X$ in the rest of this section.

\begin{figure}
\centering
\includegraphics[width=0.6\textwidth]{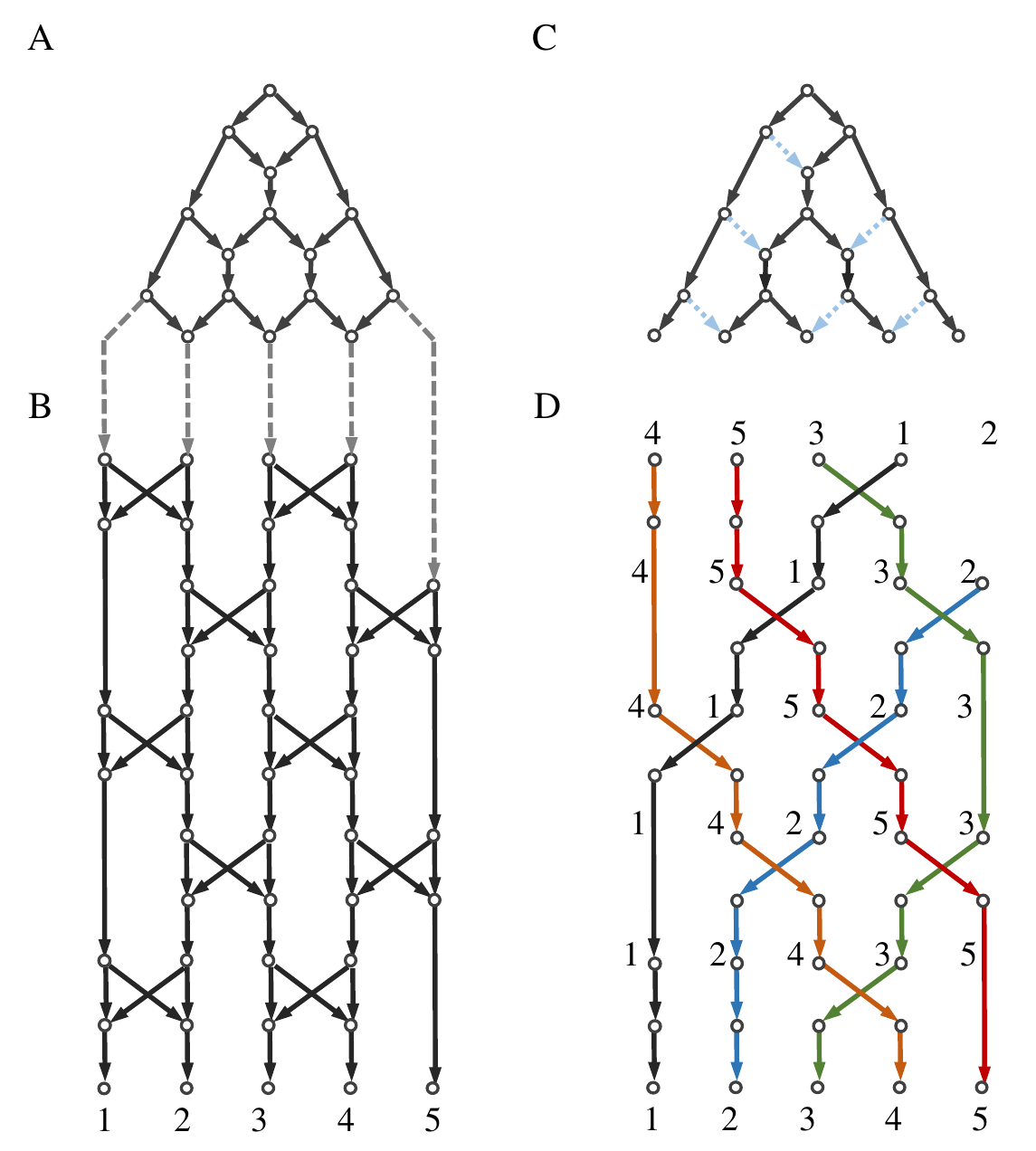}
\caption{The universal network $U$ with five leaves. (A) The upper part  $U_{\mbox{upper}}$. (B)
The lower part  $U_{\mbox{lower}}$. The square dot arrows represent edges between the two parts.
(C) The display of a rooted binary tree with five unlabeled leaves in $U_{\mbox{upper}}$, where the round dot arrows represent the removed edges. In this display, the $i$-th leaf (counted from left) in the tree is mapped to the $i$-th vertex in the last row. 
(D) The vertex-disjoint paths for the one-to-one mapping  $\pi=(45312)$, where $\pi$ maps $i$ to the $i$-th digit inside the parentheses.}
\label{fig:display_alltree}
\end{figure}

\begin{figure}[t!]
\centering
\includegraphics[width=0.5\textwidth]{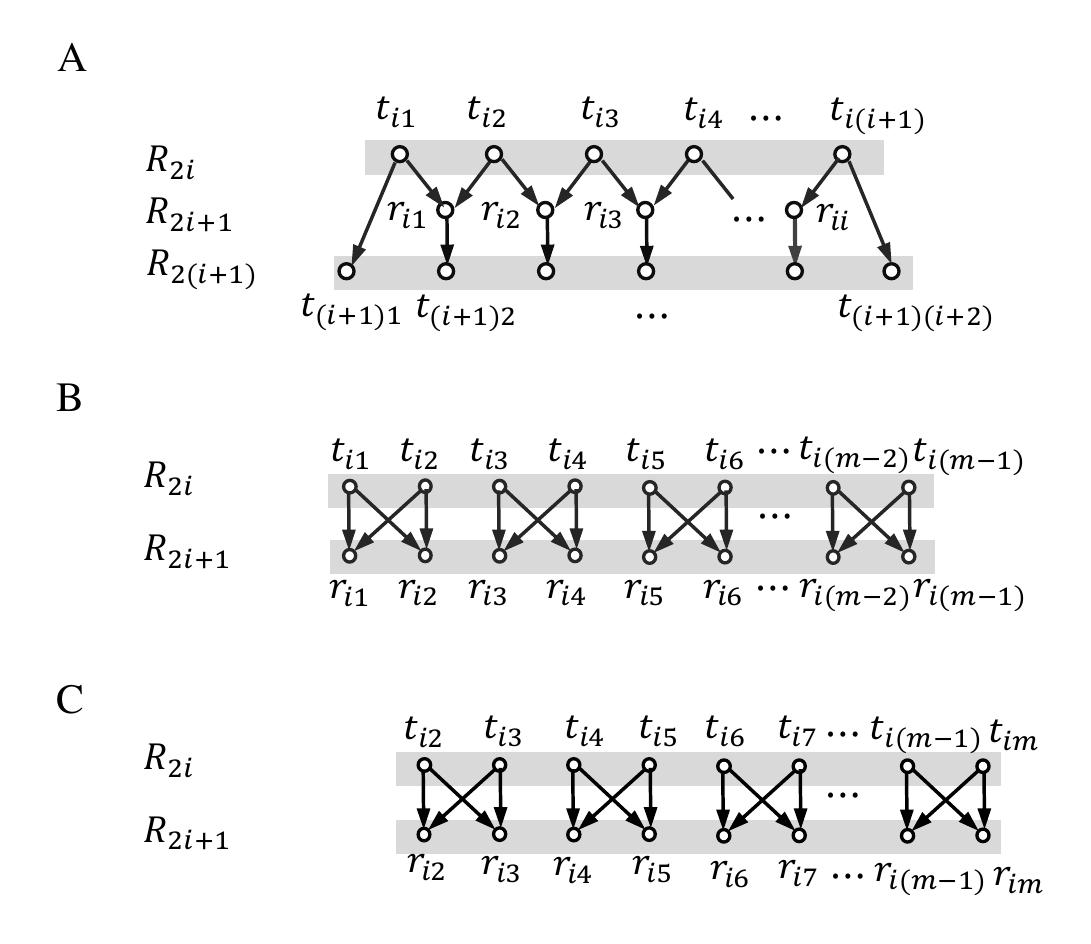}
\caption{(A)  The edges between vertexes in the rows $2i$, $2i+1$ and $2(i+1)$ in the upper part $U_{\mbox{top}}$ for each $i$. (B) The edges between vertexes in  the rows $2i$ and $2i+1$ in the lower part $U_{\mbox{top}}$ for $i=m-1, m+1, \cdots, 2m-2$. 
(C) The edges between the vertexes in  the rows $2i$ and $2i+1$ in the lower part for $i=m, m+2, \cdots, 2m-3$. Here $m$ is odd. }
\label{connection}
\end{figure}

Let $X=\{1, 2, \cdots, m\}$, $m\geq 3$. The network  $U$ on $X$  is divided into the upper and lower parts  (Figure~\ref{fig:display_alltree}A)  and  (Figure~\ref{fig:display_alltree}B).
The upper part is denoted by $U_{\mbox{upper}}$. It  is a 
$(2m-3)$-row network in which: 
\begin{itemize}
\item the root $\rho_U$ is the unique vertex  in the row 1, written $t_{01}$; 
\item the row $2i$ comprises $i+1$ tree vertexes $t_{i1}, t_{i2}, \cdots, 
t_{i(i+1)}$ for $i=1, 2, ..., m-2$; 
\item the row $2i+1$ comprises $i$ reticulation vertexes 
$r_{i1}, r_{i2}, \cdots, r_{ii}$ for $i=1, 2, ..., m-2$;
\item the edge set comprises (Figure~\ref{connection}A):
\begin{eqnarray*}
\mbox{(middle diagonal edges)}  & (t_{ij}, r_{ij}),  (t_{i(j+1)}, r_{ij}), \;\; 1\leq j\leq i,\;1\leq i\leq m-2,\\ 
\mbox{(side edges)}  &(t_{i1}, t_{(i+1)1}), \;
(t_{i(i+1)}, t_{(i+1)(i+2)}),\;  0\leq i\leq m-3,\\
\mbox{(middle vertical edges)}  & (r_{ij}, t_{(i+1)(j+1)}),\;\;  1\leq j\leq i,\;1\leq i\leq m-3.
\end{eqnarray*}
\end{itemize}

Figure~\ref{fig:display_alltree}C shows how
the rooted binary tree $(\ell_1, (((\ell_2, \ell_3), \ell_4), \ell_5))$ is displayed in $U_{\mbox{upper}}$, in which  the $i$-th leaf counted from left corresponds to $\ell_i$, $i\leq 5$.

\begin{lemma}
\label{lemma1_univ}
Let  $m\geq 3$ and let $U_{2m-4}$ consist  of the vertexes in the top $2m-4$ rows  and the edges between them in $U_{\mbox{upper}}$. Then,  every phylogenetic  tree  $T$ over $\{1, 2, \cdots, m-1\}$ is a base of $U_{2m-4}$,  where 
the $j$-th leaf {\rm (}counted from left {\rm )} in $T$ is mapped to $t_{(m-2)j}$ for each $j$ from 1 to $m-1$.
 
\end{lemma}

The lower part  $U_{\mbox{lower}}$ is essentially a rearrangeable network with $m$ inputs and $m$ outputs. A  network with $m$ inputs and $m$ outputs is said to be rearrangeable if for any one-to-one mapping $\pi$ of the inputs to the outputs, we can construct vertex-disjoint paths in the network linking the $i$th input and the 
$\pi(i)$th output for $1\leq i \leq m$
(Leighton, 1992).  
Figure~\ref{fig:display_alltree}D  shows  the vertex-disjoint paths for the mapping $\pi=(45312)$ in  $U_{\mbox{btm}}$, where $m=5$ and $\pi$ maps $i$ to 
 the $i$-th digit inside the  parentheses. 

  $U_{\mbox{lower}}$  is  a mimic of the 
rearrangeable  network derived from the well-known even-odd transposition sorting process in a linear array (Leighton, 1992, page 139). 
The topological structure of  $U_{\mbox{lower}}$ is slightly different for odd $m$ and even $m$. We use $R_i$ to denote the row $i$ in $U$.

When $m$ is odd,   $U_{\mbox{lower}}$ is divided into
$2m+1$ rows $R_{2m-2}, R_{2m-1}, \cdots, R_{4m-2}$.  
\begin{itemize}
\item For each $i=m-1, m+1, \cdots , 2m-2$, $R_{2i}$ 
comprises  $m-1$ tree vertexes $t_{ij}$ ($1\leq j\leq m-1$); $R_{2i+1}$ comprises $m-1$ reticulation vertexes 
$r_{ij}$ ($1\leq j\leq m-1$).
\item For each  $i=m, m+2, \cdots , 2m-3$, $R_{2i}$
comprise $m-1$ tree vertexes $t_{ij}$ ($2\leq j\leq m$);  $R_{2i+1}$ comprises $m-1$ reticulation vertexes 
$r_{ij}$ ($2\leq j\leq m$). 
\item The last row $R_{4m-2}$ comprises $m$ leaves labelled with $j$ ($1\leq j\leq m$) from left to right. For sake of convenience, the $j$-th leaf is denoted by $t_{(2m-1)j}$.   
\end{itemize}

\noindent The edges in $U_{\mbox{lower}}$ are formally presented in Appendix A.
Briefly, 
for $i=m-1+2j$ and $0\leq j\leq (m-1)/2$, 
the $m-1$ vertexes in $R_{2i}$ and $R_{2i+1}$ 
are paired and connected in a butterfly, as shown in Figure~\ref{connection}B.

For $i=m+2j$ and  $0\leq j\leq (m-3)/2$ 
the $m-1$ vertexes in $R_{2i}$ and $R_{2i+1}$ 
are also paired and connected in a butterfly, as shown in Figure~\ref{connection}C. 

For  $0\leq j\leq (m-3)/2$, the first reticulation  vertex $r_{(m-1+2j)1}$  in 
$R_{2m+4j-1}$ is connected with the first tree vertex  $t_{(m+1+2j)1}$ in  $R_{2m+4j+1}$, whereas the $m$-th reticulation  vertex $r_{(m+2j)m}$
in $R_{2m+4j+1}$ 
 is connected with the $m$-th tree vertex $t_{(m+2+2j)m}$  in  $R_{2m+4j+4}$. 

Finally, there are also $m$ edges between the vertexes at the bottom of  $U_{\mbox{upper}}$ and the corresponding vertexes on the top in  $U_{\mbox{lower}}$,   which are represented by the square dot
arrows drawn between Figure~\ref{fig:display_alltree}A and \ref{fig:display_alltree}B. 

When $m$ is even, the structure of $U_{\mbox{lower}}$ is presented in Appendix A.


\begin{lemma}
\label{lemma2_univ}
Let 
$\pi$ be any one-to-one mapping on $\{1, 2, \cdots, m\}$.

{\rm (i)} When $m$ is odd, there are $m$ vertex-disjoint paths  connecting $t_{(m-1)j}$ and 
$t_{(2m-1)\pi(j)}$ {\rm (}$1\leq j\leq m-1${\rm )}
and $t_{mm}$ and $t_{(2m-1)\pi(m)}$ in $U_{\mbox{lower}}$.

{\rm (ii)}  When $m$ is even, there are $m$ vertex-disjoint  paths  connecting $t_{(m-1)j}$ and 
$t_{(2m-1)\pi(j)}$ {\rm (}$1\leq j\leq m${\rm )} in $U_{\mbox{lower}}$. 

\noindent Additionally, every vertex in
$U_{\mbox{lower}}$ appears in one of the $m$ paths mentioned in  {\rm (i)} and {\rm (ii)}.
\end{lemma}

\begin{theorem}
  Every  phylogenetic  tree over $X$
  is a base for  $U$. 
\end{theorem}
\noindent {\bf Proof.} Essentially, we shall prove that  for each tree $T$, its topological structure  can be displayed in $U_{\mbox{upper}}$ and the leaves are then rearranged in $U_{\mbox{lower}}$ according to the order they appear in $T$. 
We just prove the theorem for odd $m$. (The case  $m$ is even is similar.)

Consider a phylogenetic tree $T$ over 
$X=\{1, 2, \cdots, m\}$.  Assume that its leaves are listed as $\ell_{1}, \ell_2, \cdots, \ell_{m}$ from left to right in $T$, where $1\leq \ell_j\leq m$ for each $j$. Then,  
there exists $j_0$ such that $\ell_{j_0}$ and 
$\ell_{j_0+1}$ are siblings. Let  $p_0$ be their parent.  Then, 
$T-\{\ell_{j_0}, \ell_{j_0+1}\}$ has $m-1$ leaves including $p_0$. 

By Lemma~\ref{lemma1_univ}, 
$T-\{\ell_{j_0}, \ell_{j_0+1}\}$ is displayed as a base  in the first $2m-4$ rows such that (i)  $\ell_j$ ($1\leq j<j_0$) is mapped to $t_{(m-2)j}$, (ii) $p_o$ is mapped to $t_{(m-2)j_0}$, and (iii) $\ell_{j}$
($j_{0}+1 < j \leq m$) is mapped to 
$t_{(m-2)(j-1)}$. Note that all the leaves in $T-\{\ell_{j_0}, \ell_{j_0+1}\}$  are one-to-one assigned to the tree vertexes in
$R_{2m-4}$. 

The  display of $T-\{\ell_{j_0}, \ell_{j_0+1}\}$ can be extended into a display of $T$ only by
(i) reassign $\ell_j$ to $t_{(m-1)j}$ for
$j<j_0$, (ii) assign $\ell_{j_0}$ and $\ell_{j_0+1}$ to $t_{(m-1)j_0}$ and 
$t_{(m-1)(j_0+1)}$, (iii) reassign
$\ell_j$ to $t_{(m-1)j}$ for $j_{0}+1<j\leq m-1$, and (iv)  assign $\ell_{m}$ to $t_{mm}$. 
It can be verify that such a display of $T$ does not have any dummy vertex.

Define 
$\pi=\left(\ell_1 \ell_2\cdots \ell_m\right)$. 
Clearly, $\pi$ is a one-to-one mapping over $X$, which maps $i$ to $\ell_i$.
%
By Lemma 2, there are vertex-disjoint paths that cover every vertex  and connect
$t_{(m-1)j}$ and $t_{(2m-1)\ell_j}$ ($1\leq j\leq m-1$) and $t_{mm}$ to $t_{(2m-1)\ell_m}$ in $U_{\mbox{lower}}$. Combining the display of $T$ and the $m$ disjoint paths, we conclude that  $T$ is a tree base for $U$. 
$\Box$

\section{Concluding Remarks} 



The universal tree-based network we have constructed has an important implication. 
A class of phylogenetic networks is said to be
{\it complete} if every collection of phylogenetic trees on a set of species can be displayed in a phylogenetic network on the same species in the class.  
We use  $U_X$ to denote the universal tree-based network on
$X$ for a set $X$ of species. 
Since every phylogenetic tree on $X$ is displayed in $U_X$, the class of tree-based networks is complete. 

In contrast,  the class of reticulation visible networks is incomplete as well as its subclasses such as galled trees (Wang {\it et al.}, 2001) and galled networks. 
In fact, since a reticulation visible network over $X$ has at most $4(n-1)$ reticulation vertexes (Gambette et al.,  2015), more than $2^{4(n-1)}$ different trees on $X$ cannot all be displayed in a reticulation visible phylogenetic tree simultaneously. 

The completeness  suggests that tree based networks are widespread in the entire space of phylogenetic networks. The simple linear time algorithm for testing whether a phylogenetic network is tree-based or not, given here,   is definitely useful for further examination of the distribution of tree based networks. 

Finally, tree-based networks are a natural model for horizontal gene transfer.  They also  compose a large complete class. Therefore,  it is important to study how to reconstruct a tree-based network with as few reticulation vertexes as possible from a set of gene trees or from sequence data in future.

\newpage

\section{Acknowledgment}

The author would like to thank Mike Steel for useful discussion on tree based networks.
The work was financially supported by Singapore Ministry of Education Academic Research Fund MOE2014-T2-1-155.


\newpage 
\section{Appendix A}

\subsection{Proof of Lemma 2.}
Let $m\geq 3$. Recall that
$R_i$  denotes the set of vertexes in the  row $i$ in $U$.
Let $U_{2k-2}$ be the subnetwork consisting of vertexes in the top $2(k-1)$ rows  in  $U_{\mbox{upper}}$,  
$k=2, 3, \cdots,  m-2$. The leaves in $U_{2k-2}$ are all the tree vertexes $t_{(k-1)1}$, $t_{(k-1)2}, \cdots ,
t_{(k-1)k}$ in $R_{2k-2}$. Note that 
$U_{2m-4}=U_{\mbox{upper}}  -\{ r_{(m-2)j}\;|\; 1\leq j\leq  m-2\}$.

 For each $k$, a vertex in a subtree of  $U_{2k-2}$ is said to be a dummy leaf if it is not in the lowest level $R_{2k-2}$,  but has out-degree 0. We prove that every phylogenetic tree with
$k$ leaves is a base of $U_{2k-2}$ by induction on $k$.  

When $k=2$, $N_2$ is the unique binary tree with 2 leaves. Therefore, the statement is  true.  

Assume the statement is true for $k-1$.  Let
$T$ be a phylogenetic tree with $k$  leaves.

The depth of a vertex in a rooted tree is defined to be the number of edges in the path from the  root to the vertex in the tree. 
The depth of the root is set to 0. 
Let $\ell'$ be the leaf with the largest depth in $T$. The sibling $\ell''$  of $\ell'$ must be also a leaf of $T$. (If $\ell''$ were not a leaf, its children would have a greater depth than $\ell'$.)
We use $p_(\ell')$ to denote the parent of $\ell'$ and $\ell''$ in $T$.  Clearly,  $T'=T-\{\ell', \ell''\}$ is a tree with $j-1$ leaves, one of which is $p(\ell')$. 

By the induction hypothesis, $T'$ is a base of $U_{2(k-1)-2}=U_{2k-4}$.  Let $E'\subset {\cal E}(U_{2k-4})$ such that $U_{2k-4}-E'$ is a subdivision of $T'$ in which 
there is no dummy leaf and all the leaves of $T'$ correspond one-to-one the tree vertexes
 $t_{(k-2)j}$ ($1\leq j\leq k-2$) in the row 
$R_{2k-4}$.  Assume that  $p(\ell')$ corresponds to  $t_{(k-2)j_0}$ for some 
$1\leq j_0\leq k-2$. 

Define  $E''=\{ (t_{(k-2)j}, r_{(k-2)j}) \;|\;
 j<j_0\} \cup \{ (t_{(k-2)j}, r_{(k-2)(j-1)})\;|\;
     j>j_0\}$. Then, 
$N_{k}-E'-E''$ is a subdivision of $T$ in which (i) there is no dummy leaf,
(ii) the $j$th leaf corresponds to the $j$th vertex $t_{(k-1)j}$ in $R_{2k-2}$ for $j<j_0$, 
(iii)
$\ell'$ and $\ell''$ correspond to $t_{(k-1)j_0}$ and $t_{(k-1)(j_0+1)}$, 
respectively, and (iv) the $j$th leaf corresponds to the $j$th vertex $t_{(k-1)(j+1)}$ in $R_{2k-2}$ for $j\geq j_0+1$.

This concludes the proof of Lemma 2. $\Box$

\subsection{The structure of  $U_{\mbox{lower}}$}

When $m$ is odd,  $U_{\mbox{lower}}$ consists of the last $2m+1$ rows in 
$U$: $R_{2m-2}, R_{2m-1}, \cdots, R_{4m-2}$. 

For $i=m-1, m+1, \cdots , 2m-2$, $R_{2i}$ 
comprises $m-1$ tree vertexes $t_{ij}$ ($1\leq j\leq m-1$). $R_{2i+1}$ comprises  $m-1$ reticulation vertexes 
$r_{tj}$ ($1\leq j\leq m-1$). 

For $i=m, m+2, \cdots , 2m-3$, $R_{2i}$
comprises $m-1$ tree vertexes $t_{ij}$ ($2\leq j\leq m$); 
$R_{2i+1}$ comprises  $m-1$ reticulation vertexes $m-1$ 
$r_{ij}$ ($2\leq j\leq m$).  

The last row $R_{4m-2}$ comprises  $m$ leaves  each  labelled with $\ell_j$ ($1\leq j\leq m$)  from left to right. For sake of convenience,  the $j$-th leaf is denoted by $t_{(2m-1)j}$.

The edges in $U_{\mbox{lower}}$  include: 
\begin{eqnarray*}
\mbox{({\it Vertical edges})}&&(t_{ij}, r_{ij}), 
\; 1\leq j\leq m-1,\;i=m-1+2k,\;0\leq k\leq (m-1)/2;\\ 
&&(t_{ij},  r_{ij}),
\; 2\leq j\leq m, \;i=m+2k,\;0\leq k\leq (m-3)/2;\\
&& (r_{ij}, t_{(i+1)j}),
\;2\leq j\leq m-1,\; m-1\leq i\leq 2m-3;\\
&& (r_{i1}, t_{(i+2)1}),\;i=(m-1)+2k, 0\leq k\leq (m-3)/2;\\
&& (r_{im}, t_{(i+2)m}),\; i=m+2k,\;0\leq k\leq (m-3)/2; \\
&& (r_{(2m-2)1}, t_{(2m-1)1});
\end{eqnarray*}
\begin{eqnarray*}
\mbox{({\it Diagonal edges})}&&(t_{ij}, r_{i(j+1)}),\;
(t_{i(j+1)}, r_{ij}),\\
 && \;\; j=1, 3, \cdots, m-2,\;i=m-1+2k,\;
 0\leq k\leq (m-1)/2;\\
&&(t_{ij}, r_{i(j+1)}), \;
(t_{i(j+1)}, r_{ij}), \\
&& \;\; j=2, 4, \cdots, m-1,\;i=m+2k,\;0\leq k\leq (m-3)/2;\\\\
\end{eqnarray*}

\noindent 
Finally,   there are $m$ edges connecting $m$ vertexes at the bottom in $U_{\mbox{upper}}$ and the corresponding vertexes  in  $U_{\mbox{lower}}$:
\begin{eqnarray*}
&&(t_{(m-2)1}, t_{(m-1)1}), \;(t_{(m-2)(m-1)}, t_{mm}),\\
&&(r_{(m-2)j}, t_{(m-1)(j+1)}), \;j=1, 2, \cdots , m-2. 
\end{eqnarray*}

\begin{figure}[!t]
\centering
\includegraphics{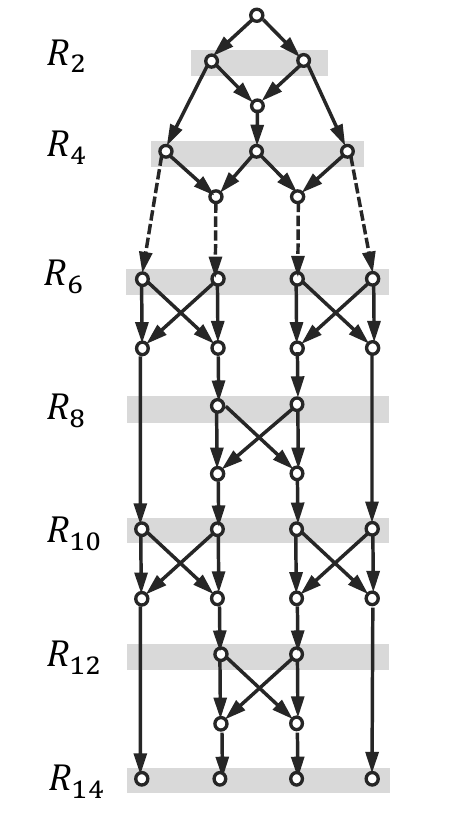}
\caption{A universal network $U$ with four leaves. Here, the square dot arrows represent the edges between the upper and lower parts. }
\label{even_case}
\end{figure}

When $m$ is even,   $N_{\mbox{lower}}$ also has $2m+1$ levels, each has $m$ or $m-2$ vertexes, as shown in Figure~\ref{even_case}.

For $i=m-1, m+1,   \cdots,  2m-3$, $R_{2i}$ comprises  $m$ tree vertexes 
  $t_{ij}$ ($1\leq j\leq m$), and $R_{2i+1}$ comprises 
   $m$ reticulation vertexes $r_{ij}$ ($1\leq j\leq m$).
  
 For $i=m, m+2, \cdots, 2m-2$,  $R_{2i}$ comprises  $m-2$ tree vertexes 
  $t_{ij}$ ($2\leq j\leq m-1$), and $R_{2i+1}$ comprises 
   $m-2$ reticulation vertexes $r_{ij}$ ($2\leq j\leq m-1$).

 The lasr row  $R_{4m-2}$ consists of  $m$ leaves with labels $\ell_j$ from left to right, 
 denoted by $t_{(2m-1)j}$ ($1\leq j\leq m$).

$U_{\mbox{lower}}$ contains the following edges:
 \begin{eqnarray*}
\mbox{({\it Vertical edges})} && (t_{ij}, r_{ij}), \;(r_{ij}, t_{(i+1)j}),\\
   &&  \;\;2\leq j\leq m-1, \; m-1\leq i\leq 2(m-1);\\
&& (r_{i1}, t_{(i+2)1}),\;(r_{im}, t_{(i+2)m}),\\
&& \;\; i=m-1+2k, 0\leq k\leq (m-2)/2; \\
%
\mbox{({\it Diagonal edges})}  && (t_{ij}, r_{i(j+1)}), \;(t_{i(j+1)}, r_{ij}),\\
 &&  
  \;\; j=1, 3, \cdots, m-1, \\
  && \;\; i=m-1+2k,\;0\leq k\leq (m-2)/2;\\
&& (t_{ij}, r_{i(j+1)}),\;(t_{i(j+1)}, r_{ij}),\\
&& 
  \;\; j=2, 4, \cdots, m-2,\\
  && \;\; i=m+2k, \;0\leq k\leq (m-2)/2.
\end{eqnarray*}

\noindent Finally,   there are $m$ edges connecting the $m$ vertexes in  $U_{\mbox{upper}}$ and the corresponding vertexes in  $U_{\mbox{lower}}$:
\begin{eqnarray*}
&&(t_{(m-2)1}, t_{(m-1)1}), \;(t_{(m-2)(m-1)}, t_{(m-1)m}),\\
&&(r_{(m-2)j}, t_{(m-1)(j+1)}), \;j=1, 2, \cdots , m-2. 
\end{eqnarray*}

\subsection{ Proof of Lemma 3.}

We first consider the case  $m$ is even. 
We define $m+1$ vectors
$S_i$ on $\{1, 2, \cdots, m\}$ ($0\leq i\leq m$).
We use $S_{i}[j]$ to denote its 
$j$-th component for $1\leq j\leq m$. 
 
Initially,  $S_{0}=(\pi(1), \pi(2), \cdots, \pi(m))$.
After  $S_i$ is computed,  we compute $S_{i+1}$ as follows.

When $i$ is even, 
$S_{i+1}$ is defined by: 
\begin{eqnarray*}
 S_{i+1}[j]=\begin{cases}
             S_i[j] & \mbox{ if } S_i[j]\leq S_i[j+1];\\
             S_{i}[j+1] &  \mbox{ if } S_i[j] > S_i[j+1]\\
            \end{cases}
\end{eqnarray*}
and 
\begin{eqnarray*}
 S_{i+1}[j+1]=\begin{cases}
             S_i[j+1] & \mbox{ if }  S_i[j]\leq S_i[j+1];\\
             S_{i}[j] &  \mbox{ if }  S_i[j] > S_i[j+1]\\
            \end{cases}
\end{eqnarray*}
for $j=1, 3, \cdots, m-1$.

When $i$ is odd,  
$S_{t+1}$ is defined by: 
\begin{eqnarray*}
  S_{i+1}[1]=S_{i}[1], \\
  S_{i+1}[m]=S_{i}[m],
\end{eqnarray*}
\begin{eqnarray*}
 S_{i+1}[j]=\begin{cases}
             S_i[j] & \mbox{ if }  S_i[j]\leq S_i[j+1];\\
             S_{i}[j+1] &  \mbox{ if }  S_i[j] >  S_i[j+1]\\
            \end{cases}
\end{eqnarray*}
and 
\begin{eqnarray*}
 S_{i+1}[j+1]=\begin{cases}
             S_i[j+1] & \mbox{ if }  S_i[j]\leq S_i[j+1];\\
             S_{i}[j] &  \mbox{ if }  S_i[j] > S_i[j+1]\\
            \end{cases}
\end{eqnarray*}
for $j=2, 4, \cdots, m-2$.

Since we emulate the odd-even transposition sorting on an array with $m$ elements (Leighton, 1992, page 129), 
$S_{m}=(\pi(1), \pi(2), \cdots, \pi (m))$. 
Using $S_{i}$'s, we obtain $m$ vertex-disjoint paths connecting $t_{(m-1)i}$ and $t_{(2m-1)\pi(i)}$
as follows:

For $k=m-1, m+1, \cdots, 2m-3$, and 
$j=1, 3, \cdots, m-1$, 
delete the vertical edges 
$(t_{kj}, r_{kj})$ and
$(t_{k(j+1)}, r_{k(j+1)})$
if $S_{k-m+2}[j]=S_{k-m+1}[j+1]$ and $S_{k-m+2}[j+1]=S_{k-m+1}[j]$; 
and delete  the diagonal edges 
$(t_{kj}, r_{k(j+1)})$ and $(t_{k(j+1)}, r_{kj})$
if $S_{k-m+2}[j]=S_{k-m+1}[j]$ and $S_{k-m+2}[j]=S_{k-m+1}[j]$.

For $k=m, m+2, \cdots, 2m-2$, and 
$j=2, 4, \cdots, m-2$,  
delete the vertical edges 
$(t_{kj}, r_{kj})$ and
$(t_{k(j+1)}, r_{k(j+1)})$
if $S_{k-m+2}[j]=S_{k-m+1}[j+1]$ and $S_{k-m+2}[j+1]=S_{k-m+1}[j]$, 
and delete the diagonal edges 
$(t_{kj}, r_{k(j+1)})$ and $(t_{k(j+1)}, r_{kj})$
if $S_{k-m+2}[j]=S_{k-m+1}[j]$ and $S_{k-m+2}[j]=S_{k-m+1}[j]$.

Since $S_{m}=(\pi (1), \pi(2), \cdots, \pi (m))$, 
the resulting $m$ vertex-disjoint paths connect 
$t_{(m-1)j}$ to $t_{(2m-1)\pi(i)}$ and pass through every vertex in $U_{\mbox{lower}}$.

For the case $m$ is odd, we can prove the statement similarly. $\Box$

\end{document}